\documentclass{Odyssey2026}

 \interspeechcameraready

\title{A Comparison of SSL-Based Feature Extractors and Back-End Classifiers for Spoofing Detection: A Multi-Corpus Training and Cross-Linguistic Analysis}

\author[affiliation={1}]{Anh-Tuan}{Dao}
\author[affiliation={1}]{Driss}{Matrouf}
\author[affiliation={1}]{Mickael}{Rouvier}
\author[affiliation={2}]{Nicholas}{Evans}

\affiliation{Laboratoire Informatique d’Avignon}{Avignon Universite}{France}
\affiliation{EURECOM }{ Sophia Antipolis}{France }
\email{\{anh-tuan.dao, driss.matrouf, mickael.rouvier\}@univ-avignon.fr, evans@eurecom.fr}
\keywords{Spoofing Detection, Multi-Corpus Training}

\usepackage{comment}
\usepackage{soul}

\usepackage{multirow}
\usepackage{xcolor}
\usepackage{subcaption}

\usepackage{numprint}
\npdecimalsign{.}

\newcommand{\dcf}[1]{\nprounddigits{2}\numprint{#1}}

\begin{document}

\maketitle

\author[affiliation={1}]{Anh-Tuan}{Dao}
\author[affiliation={1}]{Driss}{Matrouf}
\author[affiliation={1}]{Mickael}{Rouvier}
\author[affiliation={2}]{Nicholas}{Evans}


\affiliation{Laboratoire Informatique d’Avignon}{Avignon Universite}{France}
\affiliation{EURECOM }{ Sophia Antipolis}{France }
\email{\{anh-tuan.dao, driss.matrouf, mickael.rouvier\}@univ-avignon.fr, evans@eurecom.fr}

\keywords{Spoofing Detection, Multi Corpus}

\abstract{
Voice biometric systems face growing threats from spoofing attacks, yet the evaluation of detection models remains inconsistent across datasets. To investigate these unpredictable fluctuations, we conduct a comprehensive benchmark of four self-supervised learning feature extractors paired with four back-end classifiers. We compare the hierarchical local feature extraction of ResNet with the global sequence and relational modeling of attention and graph-based back-ends. Through multi-corpus training across three scenarios and six evaluation datasets, our empirical analysis yields two critical findings. First, we expose a domain bias within the ASVspoof 5 dataset, showing that naive data scaling actively degrades performance. Second, our cross-linguistic analysis reveals that fine-tuning with just 8 hours of target-language data enhances detection robustness. Together, these findings emphasize the critical need for domain-aware and language-specific adaptation in spoofing detection.


}

\section{Introduction}

Automatic Speaker Verification (ASV) offers a reliable and convenient biometric recognition method based on voice characteristics. However, advanced spoofing attacks which mimic bona fide users pose serious security risks. To counter these threats, spoofing detection systems, also called countermeasures (CMs), are increasingly deployed to protect ASV integrity. The ASVspoof~\cite{ASVspoof24} community, with global participation, has driven major advancements in this field.

Recent years have witnessed a paradigm shift in spoofing detection with the adoption of self-supervised learning (SSL) models~\cite{AASIST,conformer,MHFA_Spoof,SLS, Mamba}. By exploiting large-scale pre-training with raw speech, SSL-based models learn expressive representations which outperform conventional supervised architectures~\cite{AASIST1,RawNet2,dao24_asvspoof}. Despite these gains, robust generalization across datasets and attack types remains a challenge. Models trained using one particular corpus often exhibit sharp performance degradations when evaluation is performed with other corpora.

A conventional hypothesis in machine learning is that incorporating additional, diverse data into the training pipeline should yield a more generalized and robust model. However, in the context of spoofing detection, our investigations reveal a counter-intuitive phenomenon: expanding the training set with additional spoofing corpora frequently leads to performance degradation on unseen evaluation sets. This observation suggests the presence of hidden dataset biases that override generalized spoofing cues. To diagnose the underlying mechanics of this performance degradation, we visualized the learned representations using t-SNE. The projections reveal a critical vulnerability in multi-corpus training: alongside the expected separation of bona fide and spoofed utterances, the embeddings exhibit distinct clustering driven by their source dataset. This strong domain separation provides empirical evidence of dataset bias, demonstrating that the network could overfit to dataset-specific shortcuts rather than extracting generalizable spoofing artifacts.

Despite advances in spoofing back-end classifiers, the integration of deep convolutional architectures, specifically ResNets, with SSL-based feature extractors remains largely underexplored. Recent literature heavily favors attention-based sequence aggregators (such as MHFA and Conformer) or specialized graph-based models (like AASIST). While highly effective, attention mechanisms primarily capture global sequence dependencies, whereas graph-based models focus on relational structures. In contrast, ResNets provide a hierarchical extraction of local features. Given the complex, high-dimensional nature of the representations generated by modern SSL front-ends, we hypothesize that processing these embeddings through a robust convolutional architecture could capture localized spoofing artifacts that global aggregators might miss. To investigate this, we systematically compare ResNet against AASIST, Conformer, and MHFA. To this end, we conduct a comprehensive evaluation of multiple SSL feature extractors paired with diverse back-end classifiers. 

Our primary contributions are as follows:
\begin{itemize}
    \item \textbf{Empirical Diagnosis of Dataset Bias in Multi-Corpus Training -} Using t-SNE visualizations, we provide empirical evidence that naive multi-corpus training could causes models to rely on dataset-specific shortcuts, clustering embeddings by source domain. 
    \item \textbf{Performance Comparison of Back-end Classifiers -} We compare the performance of the ResNet back-end against three back-ends, including AASIST, Conformer, and MHFA. This comparative analysis benchmarks the empirical effectiveness of a convolutional, local-feature approach against global sequence aggregators and graph-based models when processing high-dimensional SSL representations.
    \item \textbf{Performance Comparison of Front-end Classifiers -} We isolate the impact of the front-end by evaluating four SSL feature extractors (Wav2vec2, HuBERT, WavLM, and XLSR). Our findings show that the scale and diversity of pre-training data are directly correlated with generalization, multilingual pre-training (XLSR) yielding significantly more robust representations for spoofing detection.
    \item \textbf{Cross-linguistic Evaluation -} We extend our investigation to cross-linguistic scenarios by evaluating spoofing detection performance using Spanish (HABLA dataset) and Chinese (CFAD dataset) language data. Our findings indicate that even a modest inclusion of target language training data leads to marked improvements in detection performance, underscoring the importance of language-specific adaptation.
    
\end{itemize}

\section{SSL-based Spoofing Detection}

\begin{figure}[t]
\centering
\includegraphics[width=1.0\linewidth]{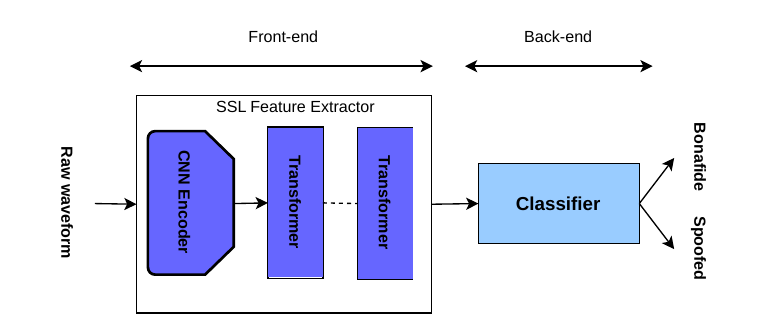}
\caption{SSL-based spoofing detection model architecture.}

\label{fig:idfe_arch}
\end{figure}

Our approach to spoofing detection leverages the strengths of SSL models to extract robust representations from raw audio data. 
Models are comprised of two main components: SSL-based feature extractors and back-end classifiers (Figure \ref{fig:idfe_arch}).

\subsection{SSL Feature Extractors}

The model comprises two primary components: a convolutional neural network (CNN) feature encoder and a Transformer-based context network. The feature encoder role is to process the raw audio input and transform it into a sequence of latent representations $z_{1:T}$. This transformation reduces the dimensionality of the input data and captures acoustic features. Following the feature encoder, the sequence of latent representations $z_{1:T}$, is input into a Transformer-based context network. The purpose of this network is to model the temporal dependencies and contextual information within the audio sequence, producing context representations $o_{1:T}$. The Transformer architecture, known for its self-attention mechanism, allows the model to capture long-range relationships in the data, which is crucial for understanding the structure of speech. In this study, we investigate four widely used SSL models for spoofing detection:


\begin{itemize}
    \item \textbf{Wav2Vec2}~\cite{baevski2020wav2vec}: consists of a convolutional feature encoder and a Transformer-based context network. The feature encoder converts raw speech into latent representations, while the Transformer refines these embeddings by capturing long-range dependencies, producing contextualized representations.

    \item \textbf{HuBERT}~\cite{hsu2021hubert}: differs from Wav2Vec2 in its training objective. Instead of contrastive learning, it employs masked prediction with discrete target representations derived from clustering.
    \item \textbf{WavLM}~\cite{chen2022wavlm}: extends Wav2Vec2 by incorporating gated relative position bias and a denoising training objective, making it more robust to noise and distortions.
    \item \textbf{XLSR}~\cite{XLS-R2022}: is a multilingual variant of Wav2Vec2, pre-trained on huge speech data from multiple languages.
\end{itemize}
Each SSL model captures speech features differently due to variations in architecture, training objectives, and pre-training datasets. By evaluating and comparing these models, we aim to identify which SSL-based features are most effective for spoofing detection.

\begin{table}[th]
    \centering
    \renewcommand{\arraystretch}{1.5} 
    \setlength{\tabcolsep}{1.5pt}
    \begin{tabular}{c|c|c}
    \hline
    \textbf{Layer name}     & \textbf{Output shape} & \textbf{Input}\\
    \hline
    Data-aug&64600 & \\
\hline
    SSL front-end&1024 x T & Wav2vec 2.0 \\
\hline
    Conv2D     & 
    32 x 1024 x T & 
    $\begin{bmatrix}
    3 \times 3, 32 
\end{bmatrix} $ \\
    \hline
    BasicBlock-1     & 
    32 x 512 x T & 
    $\begin{bmatrix}
    3 \times 3, 32 \\
    3 \times 3, 32
\end{bmatrix} \times 3$\\
    \hline
    BasicBlock-2     & 
    64 x 256 x T/2 & 
    $\begin{bmatrix}
    3 \times 3, 64 \\
    3 \times 3, 64
\end{bmatrix} \times 4$\\
    \hline
    BasicBlock-3     & 
    128 x 128 x T/4 & 
    $\begin{bmatrix}
    3 \times 3, 128 \\
    3 \times 3, 128
\end{bmatrix} \times 6$\\
    \hline
    BasicBlock-4     & 
    256 x 64 x T/8 & 
    $\begin{bmatrix}
    3 \times 3, 256 \\
    3 \times 3, 256
\end{bmatrix} \times 3$\\
    \hline
    Flatten     & 
    16384 x T/8 &\\
    \hline
    Mean Pooling    & 16384 & \\
    \hline
    Dense     & 2 &\\
    \hline
    \end{tabular}
    \caption{The architecture of our ResNet back-end classifier with SSL feature extractor.}
    \label{tab:resnet_34}
\end{table}

\subsection{Back-end Classifiers}

We evaluate and compare three back-end classifiers and our proposed ResNet back-end classifier:
\begin{itemize}
    \item \textbf{AASIST}~\cite{wave2vec2-seft-learning}: is specifically designed for spoofing detection, integrating spectro-temporal feature extraction with deep learning. It leverages convolutional and attention-based mechanisms to capture discriminative patterns in audio signals.
    \item \textbf{Conformer}~\cite{conformer}: combines convolutional and Transformer layers, effectively modeling both local and global dependencies in speech signals. This hybrid structure has shown strong performance in spoofing detection.
    \item \textbf{MHFA}~\cite{MHFA_Spoof}: Multi-Head Factorized Attentive (MHFA) pooling is an attention-based backend designed to efficiently aggregate Transformer embeddings using multi-head attention mechanisms with factorized attentive pooling. Originally developed for SSL-based speaker verification, MHFA employs a lightweight yet effective architecture to enhance feature representation.
    \item \textbf{ResNet}: We propose a ResNet-based back-end classifier which takes as input the final embedding produced by the SSL model. Our ResNet architecture consists of 34 layers with residual connections. The model utilizes a basic block architecture characterized by two 3×3 convolutional layers, with each convolution followed by a batch normalization layer and ReLU activation. Each basic block is followed by a dropout layer with a rate of 0.5 to prevent overfitting. The detailed architecture of our ResNet model is summarized in Table~\ref{tab:resnet_34}.
\end{itemize}

By evaluating these back-end classifiers, we aim to identify the most effective architecture for integrating SSL-based representations in spoofing detection tasks.

\begin{table*}[h!]
    \centering
    \renewcommand{\arraystretch}{1.0} 
    \setlength{\tabcolsep}{4.5pt}
    \caption{Performance of back-end classifiers with the XLSR feature extractor in three training cases (best results highlighted in bold, Average EER across five evaluation datasets in last column). Hidden subsets of ASVspoof21 LA and DF~\cite{ASVspoof21hid} are used for evaluation.
    }
    \label{table:case1}
    \begin{tabular}{ccccccccccccc}
        \toprule
        Training & Back-end & \multicolumn{2}{c}{Wild} & \multicolumn{2}{c}{ASV5 eval} & \multicolumn{2}{c}{ASV21 LA Hidden} & \multicolumn{2}{c}{ASV21 DF Hidden} & \multicolumn{2}{c}{Fake-Or-Real} & \multicolumn{1}{c}{Average} \\
        \cmidrule(r){3-13}
        && EER & minDCF & EER & minDCF & EER & minDCF & EER & minDCF & EER & minDCF & EER\\
        \midrule
        Case 1& AASIST & 8.72	&\dcf{0.2014}	&6.41	&\dcf{0.1854}	&11.48	&\dcf{0.2618}	&8.43	&\dcf{0.1969}	&5.91	&\dcf{0.1410}	&8.19\\
        & Conformer & 5.23	&\dcf{0.1416}	&5.19	&\dcf{0.1501}	&12.09	&\dcf{0.2914}	&9.24	&\dcf{0.2340}	&\textbf{2.10}	&\textbf{\dcf{0.0576}}	&6.77\\
        & MHFA & 4.71	&\dcf{0.1270}	& 5.56	& \dcf{0.1603}	&\textbf{10.86}	&\textbf{\dcf{0.2460}}	&8.63	&\dcf{0.1982}	&6.66	&\dcf{0.1743}	&7.28\\
        & ResNet & \textbf{3.96}	&\textbf{\dcf{0.1115}}	&\textbf{4.73}	&\textbf{\dcf{0.1368}}	&11.28	&\dcf{0.2497}	&\textbf{8.24}	&\textbf{\dcf{0.1815}}	&4.38	&\dcf{0.1150}	&\textbf{6.52}\\
        \midrule
        Case 2 & AASIST & 2.72 & \dcf{0.0732} & \textbf{10.24} & \textbf{\dcf{0.2967}} & 10.71 & \dcf{0.2784} & 8.39 & \dcf{0.2158} &0.59	&\dcf{0.0156}	&6.53	\\
        & Conformer & 4.02 & \dcf{0.1151} & 12.49 & \dcf{0.3533} & 11.07 & \dcf{0.2649} & 8.08 & \dcf{0.2024} &0.17	&\dcf{0.0046}	&7.16	\\
        & MHFA & 2.40 & \dcf{0.0618} & 11.04 & \dcf{0.3183} & 9.60 & \dcf{0.2320} & 6.71 & \dcf{0.1707} &0.26	&\dcf{0.0059}	&6.00	\\
        & ResNet & \textbf{2.02} & \textbf{\dcf{0.0515}} & 12.37 & \dcf{0.3583} & \textbf{8.21} & \textbf{\dcf{0.2084}} & \textbf{5.97} & \textbf{\dcf{0.1614}} &\textbf{0.17}	&\textbf{\dcf{0.0038}}	&\textbf{5.75}	\\
        \midrule
        Case 3& AASIST   & 1.80	&\dcf{0.0480}	&13.69	&\dcf{0.3913}	&5.72	&\dcf{0.1565}	&3.93	&\dcf{0.0994}	&0.48	&\dcf{0.0097}	&5.12	\\
        & Conformer & 1.69	&\dcf{0.0431}	&13.35	&\dcf{0.3858}	&\textbf{4.59}	&\dcf{0.1126}	&\textbf{2.57}	&\textbf{\dcf{0.0625}}	&0.26	&\dcf{0.0067}	& 4.49	\\
        & MHFA     & 1.45	&\dcf{0.0334}	&12.06 	&\dcf{0.3484} 	&5.86	&\dcf{0.1256}	&3.53	&\dcf{0.0742}	&0.26	&\dcf{0.0067}	&4.63	\\
        & ResNet    &\textbf{1.21} 	&\textbf{\dcf{0.0312}}	&\textbf{11.46}	&\textbf{\dcf{0.3249}}	&5.21	&\textbf{\dcf{0.1113}}	&3.03	&\dcf{0.0659}	&\textbf{0.13}	&\textbf{\dcf{0.0034}}	&\textbf{4.20}	\\
        \bottomrule
    \end{tabular}
\end{table*}

\begin{table*}[h!]
    \centering
    \renewcommand{\arraystretch}{1.0} 
    \setlength{\tabcolsep}{4.5pt}
    \caption{Performance Comparison of SSL feature extractors with our ResNet back-end in training case 2. }
    \label{table:ssl-based}
    \begin{tabular}{ccccccccccccc}
    \toprule
    Extractor& \multicolumn{2}{c}{Wild} & \multicolumn{2}{c}{ASV5 eval} & \multicolumn{2}{c}{ASV21 LA Hidden} & \multicolumn{2}{c}{ASV21 DF Hidden} & \multicolumn{2}{c}{Fake-Or-Real} & \multicolumn{1}{c}{Average} \\
    \cmidrule(r){2-3} \cmidrule(r){4-5} \cmidrule(r){6-7} \cmidrule(r){8-9} \cmidrule(r){10-11} \cmidrule(r){12-12}
     & EER & minDCF & EER & minDCF & EER & minDCF & EER & minDCF & EER & minDCF & EER \\
    \midrule

    Wav2vec2  & 9.03 & \dcf{0.2563} & \textbf{7.41} & \textbf{\dcf{0.2113}} &19.49 &\dcf{0.4969} &15.52 &\dcf{0.4110} &1.67 &\dcf{0.0483} &10.62 \\
    HuBERT  & 15.13 & \dcf{0.4193} & {9.36} & {\dcf{0.2690}} & 33.47 & \dcf{0.7432} & 29.39 & \dcf{0.6436} & 3.54 & \dcf{0.0941} & 18.18 \\
    WavLM  & 7.82 & \dcf{0.1691} & 10.12 & \dcf{0.2933} & 14.74 & \dcf{0.3578} & 10.26 & \dcf{0.2457} & 0.39 & \dcf{0.0105} & 8.67 \\
    XLSR  & \textbf{2.02} & \textbf{\dcf{0.0515}} & 12.37 & \dcf{0.3583} & \textbf{8.21} & \textbf{\dcf{0.2084}} & \textbf{5.97} & \textbf{\dcf{0.1614}} & \textbf{0.17} & \textbf{\dcf{0.0038}} & \textbf{5.75} \\
    \bottomrule
    \end{tabular}

\end{table*}

\section{Experimental Setup}
\subsection{Datasets and Metrics}\label{sec:datasets-training}
For training, we use the ASVspoof 5 training set, MLAAD-v3~\cite{mlaad2024}, ASVspoof19~\cite{ASVspoof19}, and VCTK~\cite{vctk}. To support multi-corpus evaluation, the ASVspoof 5 validation set is replaced with the ITW dataset. We define three distinct training scenarios:
\begin{itemize}
    \item Training Case 1: ASVspoof 5 training set only.
    \item Training Case 2: ASVspoof 5 training and MLAAD-v3.
    \item Training Case 3: ASVspoof 5 training set, MLAAD-v3, ASVspoof19, and VCTK.
\end{itemize}

For evaluation, we use six evaluation datasets: four English (ASVspoof21 LA and DF Hidden~\cite{ASVspoof21hid, ASVspoof21}, ASVspoof 5 evaluation set, Fake-Or-Real~\cite{FoR2019}) and two non-English (HABLA~\cite{habla}, CFAD~\cite{cfad} noisy-unseen-test). The ASVspoof 2021 LA and DF hidden subsets have non-speech segments removed, preventing models from exploiting such shortcuts. Prior work~\cite{Martin-DonasARG24} shows the challenges presented by these hidden subsets, compared to standard evaluation subsets, since they demand a reliance on core spoofing cues rather than non-speech-related shortcuts. Performance is measured using Equal Error Rate (EER) and minimum Detection Cost Function (minDCF).


\begin{figure*}[t]
    \centering
    \begin{subfigure}[b]{0.32\linewidth}
        \centering
        \includegraphics[width=1.0\linewidth]{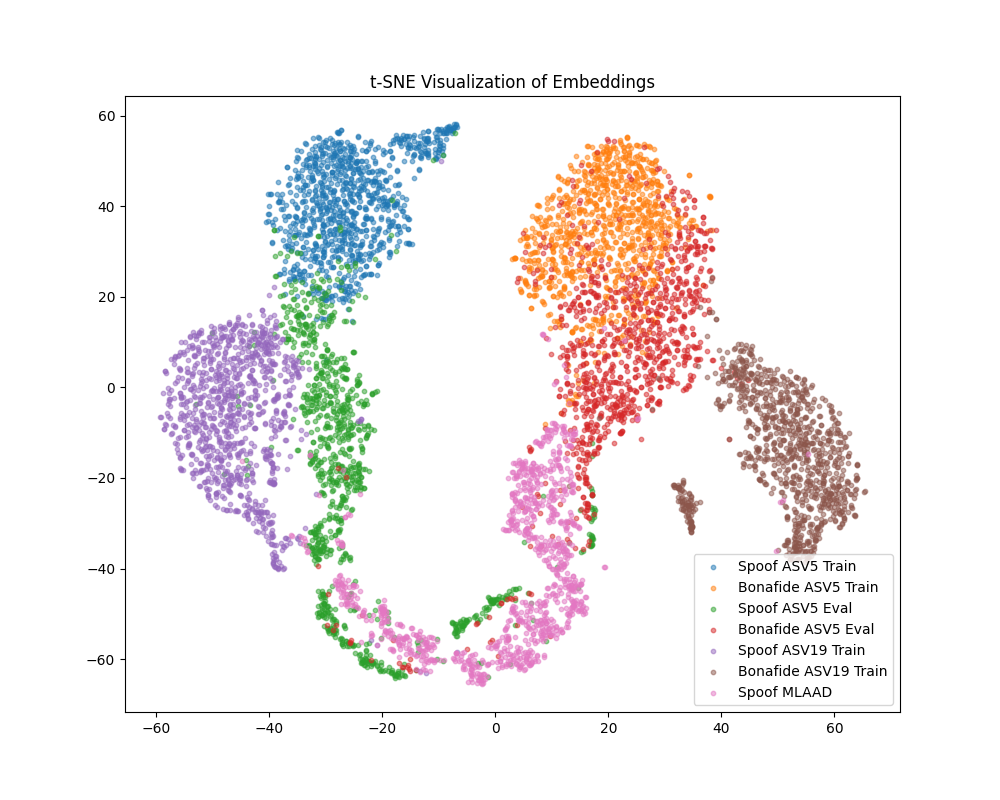}
        \caption{Training Case 1.}
    \end{subfigure}
    \hfill
    \begin{subfigure}[b]{0.32\linewidth}
        \centering
        \includegraphics[width=1.0\linewidth]{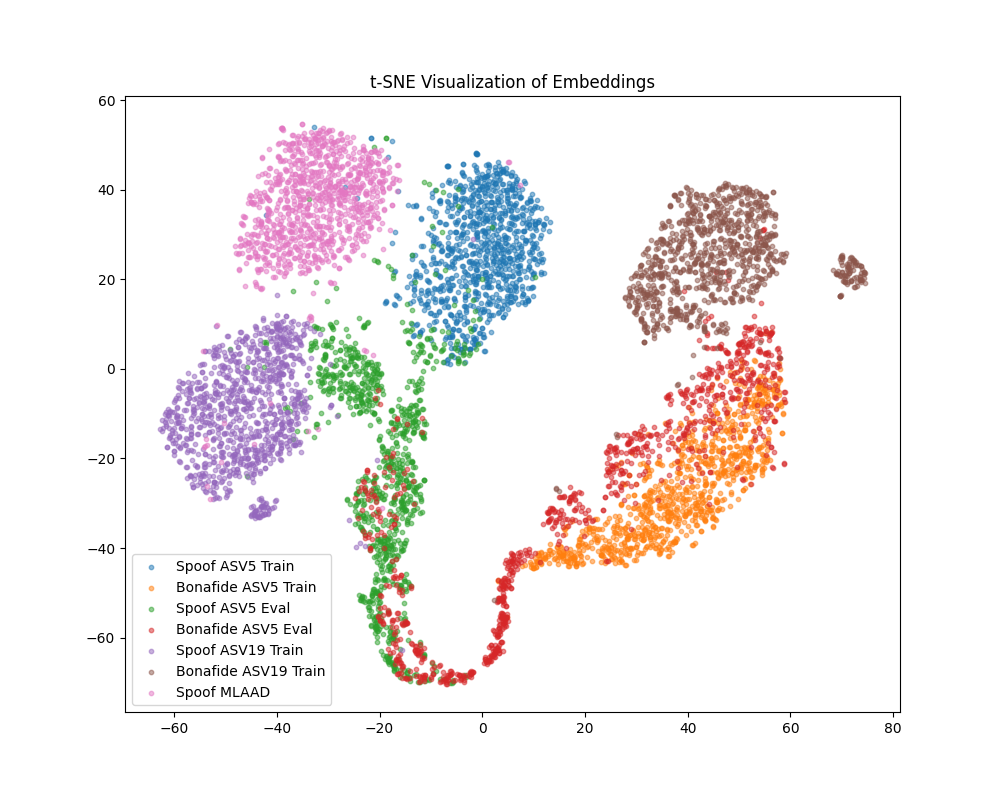}
        \caption{Training Case 2.}
    \end{subfigure}
    \hfill
    \begin{subfigure}[b]{0.32\linewidth}
        \centering
        \includegraphics[width=1.0\linewidth]{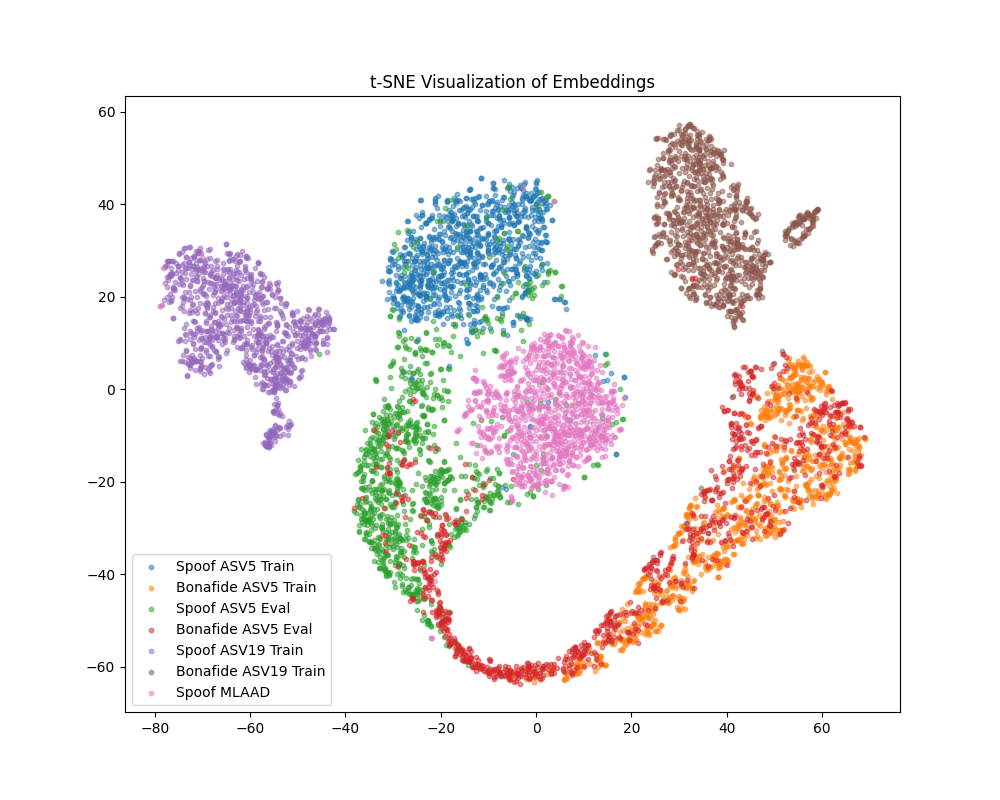}
        \caption{Training Case 3.}
    \end{subfigure}    
    \caption{t-SNE visualization of learned embeddings of XLSR-ResNet model in three training cases.}
    \label{fig:learned_emb}
\end{figure*}

\begin{table*}[h!]
\centering
\renewcommand{\arraystretch}{1.0} 
\setlength{\tabcolsep}{2.5pt}
\caption{Model performance (\% EER) of the XLSR-ResNet model in {training case 1} and case 2, evaluated on ASVspoof 5 eval break down: codecs (C01-11). The third column ("-") represents performance on non-codec audio, while "pooled" indicates the average performance across all categories.
}
\label{table:codec-case1}
\begin{tabular}{cccccccccccccc}
\toprule
       & pooled &   -    &  C01   &  C02   &  C03   &  C04   &  C05   &  C06   &  C07   &  C08   &  C09   &  C10   &  C11  \\ 
\midrule
Training Case 1 &  4.72 &  0.95 &  2.70 &  3.22 &  3.15 &  11.87 &  1.03 &  1.33 &  14.52 &  5.24 &  3.26 &  4.30 &  1.27\\ 
Training Case 2 & 12.34 &  0.53 &  8.58 &  13.68 &  6.95 &  27.11 &  3.33 &  2.52 &  32.32 &  10.24 &  9.77 &  11.66 &  1.07\\

\bottomrule
\end{tabular}
\end{table*}

\subsection{Data Augmentation}
We apply standard data augmentation techniques during training, utilizing the MUSAN corpus and the real room impulse response (RIR) database~\cite{MUSAN,Reverb2017}. Each training utterance underwent four augmentation methods:

\begin{itemize}
    \item Reverberation: Utterances are convolved with real RIRs to simulate reverberation effects associated with propagation in various acoustic spaces.
    \item Speech: A summation of three to eight different-speaker utterances are added to each training utterance at signal-to-noise ratios (SNRs) of 13-20~dB.
    \item Music: Randomly-selected music recordings from MUSAN are added to each training utterance at SNRs of 5-15~dB.
    \item Noise: Randomly-selected noise recordings from MUSAN are added to each training utterance at SNRs of 0-15~dB.
\end{itemize}

\subsection{Implementation Details}\label{sec:implem_details}
All models, including the SSL front-ends and back-end classifiers, are fine-tuned end-to-end.
During training, utterances are randomly segmented into 4-second clips before being fed into the models, whereas evaluation is performed on the complete audio clips. We employ the standard Adam~\cite{adam} optimizer with a learning rate of $10^{-6}$, a weight decay of $10^{-5}$, in order to minimize a weighted cross-entropy loss function. Training is performed in batches of 32 samples over 30 epochs using Nvidia A100 GPUs, with the best checkpoint selected.

We evaluate four SSL feature extractors from the Torchaudio library: Wav2Vec2-Large-LV60K, HuBERT-Large, WavLM-Large, and Wav2Vec2-XLSR-300m. These models consist of a CNN encoder followed by 24 Transformer layers, producing a 1024-dimensional output representation, with a total of 316 million parameters. For back-end classification, we compare our ResNet-based model against three architectures: AASIST~\cite{wave2vec2-seft-learning}, Conformer~\cite{conformer}, and MHFA~\cite{MHFA_Spoof}.

\section{Results and Analysis}

Our experimental framework systematically benchmarks the combination between various Self-Supervised Learning (SSL) front-ends and back-end classifiers. We begin by evaluating four distinct back-ends paired with an XLSR feature extractor (Sec. 4.1), followed by qualitative results of the learned embedding space via visualization (Sec. 4.2). Subsequently, we assess the compatibility of alternative SSL architectures with our ResNet back-end (Sec. 4.3). The study concludes with an investigation into the model’s cross-linguistic transferability and generalization (Sec. 4.4).

\subsection{Comparison of Back-end Classifiers}\label{section:compare-backend}
We benchmark four back-end architectures (i.e. AASIST, Conformer, MHFA, and our proposed ResNet) across three distinct training cases (see Sec. \ref{sec:datasets-training}). As summarized in Table~\ref{table:case1}, the ResNet back-end gives the best average performance, surpassing the competitive MHFA baseline with relative EER reductions of 10.4\%, 4.1\%, and 9.2\% across Cases 1, 2, and 3, respectively.

While multi-corpus training generally scales model efficacy, the results reveal a nuanced sensitivity to data composition. On the Wild dataset, ResNet’s EER decreases by 48\% (Case 2) and 69\% (Case 3) relative to Case 1. Similarly, the inclusion of the MLAAD dataset (Case 2) precipitously reduces EER from 4.38\% to 0.17\% for the Fake-or-Real evaluation set.
Conversely, a notable performance regression occurs on the ASVspoof 5 evaluation set following the introduction of external training data, where the EER increases from 4.73\% (Case 1) to 12.37\% (Case 2). To investigate this anomaly, we decouple the results by audio type (Table~\ref{table:codec-case1}). The degradation is isolated to codec-processed audio (compressed/transmitted), whereas performance on non-codec audio actually improves (EER declines from 0.95\% to 0.53\%). This suggests a significant domain mismatch or dataset bias between ASVspoof 5 and MLAAD with respect to channel effects.

\subsection{Visualisation}
To further diagnose the bias issue in the previous section, we visualize the ResNet embeddings using t-SNE (Fig.~\ref{fig:learned_emb}). In Case 1 (trained solely on ASVspoof 5), spoofed evaluation samples (green) exhibit a fragmented, non-compact distribution, with several samples encroaching on the bonafide region (red). In Case 2, while the addition of MLAAD data improves the alignment of MLAAD spoof samples with the ASVspoof 5 spoof cluster, it simultaneously pulls ASVspoof 5 bonafide embeddings toward the spoof region.
This shift introduces representation confusion, explaining the performance drop in Case 2. Furthermore, the distinct clustering by dataset source, rather than just class label, confirms the presence of dataset bias.

\subsection{Comparison of SSL-based Feature Extractors}
We evaluate four SSL front-ends integrated with our ResNet back-end (Table~\ref{table:ssl-based}). XLSR demonstrates superior efficacy, achieving a benchmark EER of 2.02\% on the Wild dataset. This robustness is attributed to its extensive pre-training on 436,000 hours of diverse, multilingual audio, which facilitates a highly generalized representation of global speech variability.
Although Wav2vec 2.0 and HuBERT exhibit higher average EERs compared to WavLM and XLSR, they remain remarkably competitive on the ASVspoof 5 evaluation set. We hypothesize that this contextualized success stems from a pre-training bias: both models were trained on the Libri-Light corpus, which was also utilized to develop several ASVspoof 5 attack models.

\begin{table}[h!]
    \centering
    \renewcommand{\arraystretch}{1.0} 
    \setlength{\tabcolsep}{1.5pt}
    \caption{Performance on non-target languages (HABLA: Spanish dataset; CFAD: Chinese dataset) in training case 1. }
    \label{table:language-case1}
    \begin{tabular}{ccccccccccccc}
    \toprule
    Method & \multicolumn{2}{c}{HABLA} & \multicolumn{2}{c}{CFAD} \\
    \cmidrule(r){2-3} \cmidrule(r){4-5} 
    & EER & minDCF & EER & minDCF\\
    \midrule
    XLSR + AASIST & 18.43 & \dcf{0.3269} & 30.16 & \dcf{0.6623} \\
    XLSR + Conformer & 13.58 & \dcf{0.2767} & 23.15 & \dcf{0.5461} \\
    XLSR + MHFA & 19.14 & \dcf{0.3424} & 22.14 & \dcf{0.4762} \\
    XLSR + ResNet & 17.90 & \dcf{0.3249} & 27.23 & \dcf{0.5589} \\
    \bottomrule
    \end{tabular}

\end{table}

\begin{table}[h!]
    \centering
    \renewcommand{\arraystretch}{1.0} 
    \setlength{\tabcolsep}{1.5pt}
    \caption{Performance on other languages in training case 2: ASVspoof 5 training and MLAAD-v3 dataset (MLAAD-v3 includes spoofed Spanish audio but excludes Chinese)
    }
    \label{table:language-case2}
    \begin{tabular}{ccccccccccccc}
    \toprule
    Method & \multicolumn{2}{c}{HABLA} & \multicolumn{2}{c}{CFAD} \\
    \cmidrule(r){2-3} \cmidrule(r){4-5} 
    & EER & minDCF & EER & minDCF \\
    \midrule
    XLSR + AASIST & 6.33 & \dcf{0.1603} & 26.88 & \dcf{0.6439} \\
    XLSR + Conformer & 4.39 & \dcf{0.1147} & 27.11 & \dcf{0.4926} \\
    XLSR + MHFA & 5.53 & \dcf{0.1283} & 23.23 & \dcf{0.4740} \\
    XLSR + ResNet & 2.98 & \dcf{0.0708} & 23.54 & \dcf{0.4560} \\
    \bottomrule
    \end{tabular}

\end{table}

\subsection{Cross-linguistic Generalization}
We evaluate the cross-linguistic robustness of our models using the Spanish (HABLA) and Chinese (CFAD) datasets (Tables~\ref{table:language-case1} and \ref{table:language-case2}).
In Case 1, under monolingual supervision on English data (ASVspoof 5), models exhibit a substantial performance deficit on non-English evaluation sets. The Conformer architecture yields the most competitive baseline, with EERs of 13.58\% on HABLA and 23.15\% on CFAD. This underscores that models optimized exclusively for English feature spaces fail to encapsulate the phonetic and prosodic signatures of spoofing attacks in diverse linguistic contexts, manifesting a significant linguistic domain gap.
The introduction of the MLAAD dataset (Case 2), which comprises 23 languages including Spanish but excluding Chinese, yielded divergent results.
On the Spanish HABLA set, EERs decline precipitously. Despite the sparsity of Spanish spoofed audio in MLAAD (approximately 8 hours), our ResNet achieves a 2.98\% EER. This demonstrates that even minimal exposure to target-language variance facilitates effective cross-linguistic alignment.
Conversely, performance on the Chinese CFAD set remains invariant.
While our results highlight the necessity of including target-language data during training, further experiments across a broader range of languages are required to generalize this hypothesis.

\section{Conclusions}

In this study, we conduct a comprehensive evaluation of spoofing detection models, focusing on the integration of SSL-based feature extractors and back-end classifiers across various training and evaluation scenarios. Our findings show the ResNet classifier to be the most effective. It achieves consistently the lowest average EER. 
Our analysis of SSL-based feature extractors reveals substantial performance variations, with XLSR standing out as the most effective due to extensive pre-training using diverse datasets. This underscores the critical role of large-scale, diverse pre-training in enhancing feature representations for the spoofing detection task. Additionally, our multi-corpus training experiments expose significant dataset biases, particularly for the ASVspoof~5 evaluation set, which highlights the need for robust bias mitigation strategies to improve generalization.
Furthermore, our cross-linguistic evaluation for Spanish (HABLA dataset) and Chinese (CFAD dataset) languages demonstrates that the incorporation of even a modest amount of target-language data (approximately 8 hours) yields marked improvements in detection performance. This finding emphasizes the importance of language-specific adaptation, particularly for underrepresented languages, to spoofing detection robustness in multilingual contexts.

\section{Acknowledgements}
This work was performed using HPC resources from GENCI-IDRIS. This work was financially supported by ANR BRUEL (ANR-22-CE39-0009).

\bibliographystyle{IEEEtran}
\bibliography{Odyssey2026_BibEntries}

\end{document}